\documentclass[a4paper,conference]{IEEEtran}
\IEEEoverridecommandlockouts

\usepackage{cite}
\usepackage{amsmath,amssymb,amsfonts}
\usepackage{algorithmic}
\usepackage{graphicx}
\usepackage{textcomp}
\usepackage{xcolor}
\usepackage{gensymb}
\usepackage{authblk}
\usepackage{hyperref}
\usepackage[nolist,nohyperlinks]{acronym}
\begin{acronym}
    \acro{mmW}{millimeter wave}
    \acro{DAC}{digital to analog converter}
    \acro{ADC}{analog to digital converter}
    \acro{BER}{bit error rate}
    \acro{RRC}{root-raised cosine}
    \acro{DUC}{digital up-converter}
    \acro{DDC}{digital down-converter}
    \acro{NCO}{numerically controlled oscillator}
    \acro{Tx}{transmitter}
    \acro{Rx}{receiver}
    \acro{DMA}{direct memory access}
    \acro{IF}{intermediate frequency}
    \acro{MSPS}{mega samples per second}
\end{acronym}

\def\BibTeX{{\rm B\kern-.05em{\sc i\kern-.025em b}\kern-.08em
    T\kern-.1667em\lower.7ex\hbox{E}\kern-.125emX}}
\begin{document}

\title{Demo: An RFSoC-Based Testbed for Over-the-Air Wireless Transceiver at Millimeter Wave Frequency}

\author{Jeet Tekchandani}
\author{Jai Mangal}
\author{Sumit J. Darak}
\author{Shobha Sundar Ram}
\affil{Electronics \& Communication Department, Indraprastha Institute of Information Technology, Delhi, India-110020}
\affil{(jeet20208, jaim, sumit, shobha)@iiitd.ac.in}


\maketitle

\begin{abstract}
The \ac{mmW} frequency spectrum has been explored recently for large bandwidth communication. At these frequencies, narrow directional beams are required for communication since the signal attenuation is high due to atmospheric absorption. This work presents an AMD RFSoC and Sivers Semiconductors analog front-end based hardware testbed capable of directional communication via analog beamforming at \ac{mmW}. The proposed testbed comprises orthogonal frequency division multiplexing (OFDM) based baseband physical layer and digital front-end on an ARM processor and field programmable gate array (FPGA), respectively, integrated with high-speed data converters of the RFSoC. The RFSoC output at sub-6GHz is integrated with a \ac{mmW} multi-antenna analog-front end for over-the-air communication at 29.8 GHz. We demonstrate end-to-end communication over the air and present \ac{BER} analysis in the presence of radio frequency impairments and beam misalignments in real radio channels. 
\end{abstract}


\section{Introduction}
Limited spectrum availability at sub-6 GHz and large bandwidth requirements for next-generation wireless applications such as vehicular networks and immersive communications, including extended reality and holographic communication, has led to the exploration of \ac{mmW} spectrum  \cite{sharma2024low,zhang2021performance}. Since \ac{mmW} frequencies exhibit higher attenuation due to atmospheric absorption, communication is carried out through beamforming-based narrow directional beams \cite{sneh2023ieee,tewari2024reconfigurable,tewari2022reconfigurable}. In this paper, we present a hardware testbed for end-to-end orthogonal frequency division multiplexing (OFDM) communication at \ac{mmW} frequencies, utilizing analog beamforming. We demonstrate end-to-end communication over the air and \ac{BER} analysis in the presence of radio frequency impairments and beam misalignments in real radio channels.


\section{Hardware Testbed}
\begin{figure}[htbp]
\centering
\centerline{\includegraphics[scale=0.7]{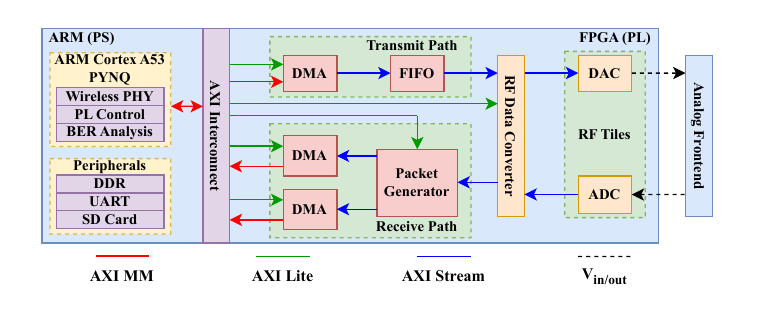}}
\caption{Hardware software co-design based baseband and digital front-end architecture of the proposed testbed.}
\label{fig:Arch}
\end{figure}

This section presents the design details of the proposed testbed consisting of AMD RFSoC 4x2 comprising of 64-bit Quad ARM Cortex™-A53 processor, ultra-scale field programmable gate array (FPGA), four 14-bit 5 Giga sample per second (GSPS) analog-to-digital converters (ADC), and two 14-bit 9.85 GSPS RF-DAC. The proposed testbed consists of an OFDM baseband physical layer for the transmitter and receiver realized on the ARM processor running the PYNQ operating system. Further, the digital front-end comprises digital up and down converters in the FPGA communicating with the ARM processor via advanced extensible interface (AXI), and single digital to analog converter (DAC) and analog to digital converter (ADC) integrated with the FPGA as shown in Fig.~\ref{fig:Arch}. 

\begin{figure}[!t]
\centerline{\includegraphics[scale=0.3]{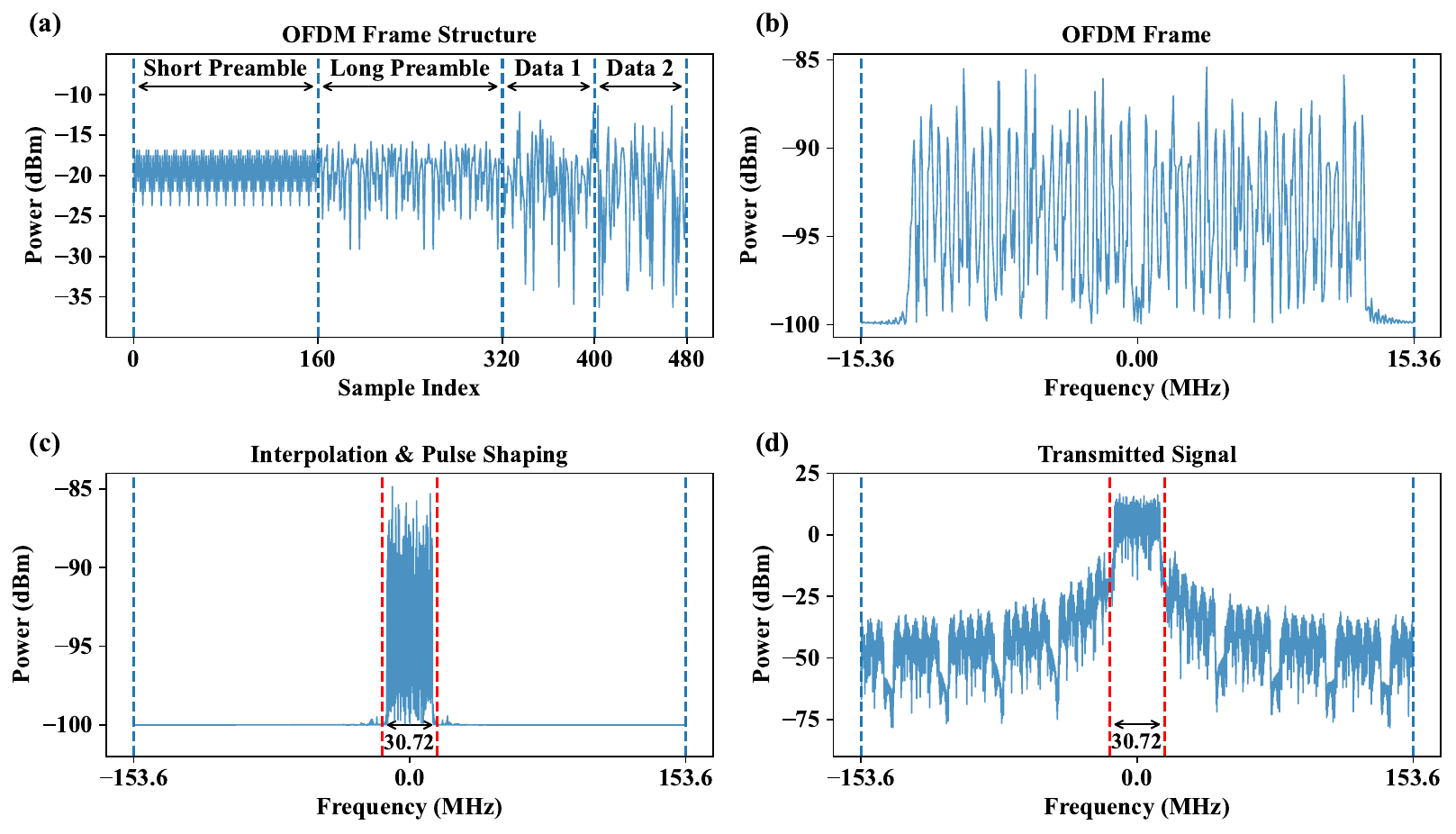}}
\caption{Transmitted OFDM signal (a) frame structure, (b) Baseband signal, (c) Digital front-end output, and (d) Input to DAC.}
\label{fig:Tx}
\end{figure}

We begin with the OFDM baseband physical layer. The OFDM frame consists of 480 complex samples, comprising four primary sections: the short preamble, the long preamble, and two data payloads. The short preamble, composed of 16-sample short training symbols repeated 10 times, is intended for coarse frequency offset estimation and timing synchronization at the receiver. Its repeated structure allows autocorrelation-based detection, which helps the receiver reliably detect the start of a packet and align frames. The long preamble is used for channel estimation and fine frequency offset correction; it consists of a 64-sample long training symbol repeated twice, preceded by a 32-sample cyclic prefix. The frame contains two data payloads, each having 64 samples and a cyclic prefix of 16 samples, there are 48 QPSK modulated data samples, 4 pilot samples, and 12 virtual samples. The OFDM frame structure in the time and frequency domain is presented in Figs.~\ref{fig:Tx}(a) and (b), respectively. 

The baseband signal, with a bandwidth of 30.72 MHz, is passed through the digital front-end. The digital front-end interpolates the signal by 10 and applies a \ac{RRC} filter that performs pulse shaping as shown in Fig.~\ref{fig:Tx}(c). The upsampled baseband signal has a sampling frequency of 307.2 MHz. We decompose the 32-bit complex samples into their 16-bit real (I) and 16-bit imaginary (Q) parts and interleave them as required by the DAC. We scale the signal to adjust its amplitude for transmission as per the dynamic range of the DAC. To do so, we normalize the data to get it within the range from -1 to +1 and scale it from $-2^{15}$ to $2^{15}-1$. The scaled signal, shown in Fig.~\ref{fig:Tx}(d), is passed to \ac{DAC} via FPGA.

The ARM processor forwards the upconverted baseband OFDM frame to the AXI \ac{DMA}, which continuously transmits the signal at a data rate of 307.2 \ac{MSPS} to the \ac{DAC} tile of the RFSoC 4x2. The DAC tile further interpolates the signal by 16, raising the sampling rate to 4.9152 GSPS. Since the \ac{mmW} analog front-end requires the carrier frequency of the input signal to be between 3.5 - 4 GHz in \ac{IF} mode, we program the \ac{NCO} available in the DAC tile to modulate the OFDM signal from the baseband to \ac{IF}  of 3.8 GHz.

\begin{figure}[htbp]
\centerline{\includegraphics[scale=0.16]{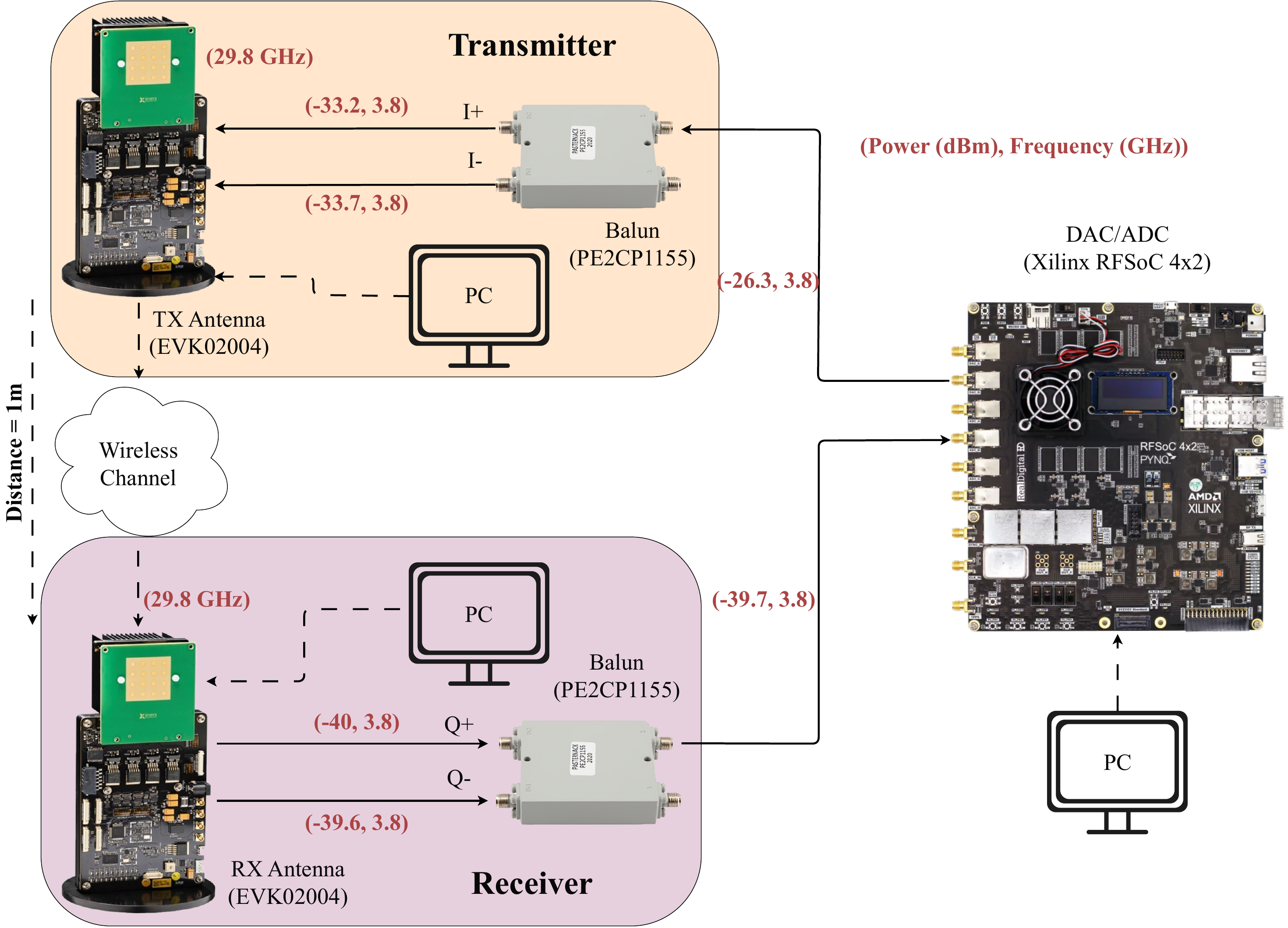}}
\caption{System level layout of the proposed testbed. The numbers within parentheses $()$ indicate (Power(dBm), Frequency(GHz)).}
\label{fig:Hardware_Arch}
\end{figure}

The output signal from the RFSoC 4x2 DAC \textcolor{black}{is at a power level of -26.3 dBm and at an \ac{IF} of 3.8 GHz.} The signal is then passed to a Pasternack PE2CP1155 balun, which converts the single-ended signal into a differential-ended signal. These differential signals enhance the transmission's noise immunity by canceling out common-mode noise and supporting balanced transmission over the wireless channel. \textcolor{black}{The balun incurs a loss of 1.5 dB. The output signal power level reduces to -33.2 dBm at I+ port and -33.7 dBm at I- port at the \ac{IF} of 3.8 GHz.} The output signal from the balun is passed to the transmitter antenna of the transceiver system from Sivers Semiconductor, EVK02004. Here, the modulator upconverts the signal from the \ac{IF} to \ac{mmW} frequency of 29.8 GHz. Additionally, it performs analog beamforming using phase shifters and a 4x4 uniform rectangular antenna array. The unit can perform analog beamforming to steer the beam from $-40\degree$ to $+40\degree$ in azimuth and from $-30\degree$ to $+30\degree$ in elevation with a step size of $10\degree$ towards the receiver on the fly. The system-level layout of the proposed testbed and link budget analysis are presented in Fig.~\ref{fig:Hardware_Arch}.

The signal is received at an EVK02004 receiver antenna array 1 m away from the transmitter. Beam-combining takes place at the receiver, and the signal is downconverted back to \ac{IF} of 3.8 GHz. The antenna outputs two differential signals, with power levels of -40 dBm from the Q+ port and 39.6 dBm from the Q- port which are passed through the balun to obtain a single-ended output signal. This signal is fed to the \ac{ADC} of the RFSoC 4x2, which samples the signal at a rate of 4.9152 GSPS. The \ac{NCO} in the \ac{ADC} tile demodulates the signal back to the baseband, generating two real streams of samples corresponding to the I and Q parts of a complex signal. The \ac{ADC} then performs a decimation by 16 and outputs the sampled data at a rate of 307.2 \ac{MSPS}. The packet generator block forwards a requested amount of samples from the two ADC output streams to the processor via the DMA blocks at the same data rate.

Once the data is received by the ARM processor, the OFDM receiver baseband physical layer performs matched filtering using an identical RRC filter as the transmitter. Packet detection and timing synchronization are performed by correlating the filtered signal with a delayed version of itself. The short preamble creates distinct correlation peaks that allow the receiver to detect the start of a packet and extract the received frame. The detected frame is decimated by 10 to restore the original frame length of 480 samples. After the frame is reconstructed, we perform coarse and fine frequency offset correction using the short and long preambles respectively. This is followed by channel estimation via least squares. The long preamble symbols known to us are used to calculate the channel response across the samples. Finally, we perform channel equalization, which corrects any amplitude and phase distortions introduced by the channel, ensuring reliable demodulation of the data payloads. After equalization, the pilot and virtual samples are removed, and the remaining samples carrying the data are QPSK-demodulated. The received and transmitted data payloads are compared to compute the \ac{BER}, our primary performance metric.

\section{Results \& Discussions}
\begin{figure}[htbp]
\centerline{\includegraphics[scale=0.15]{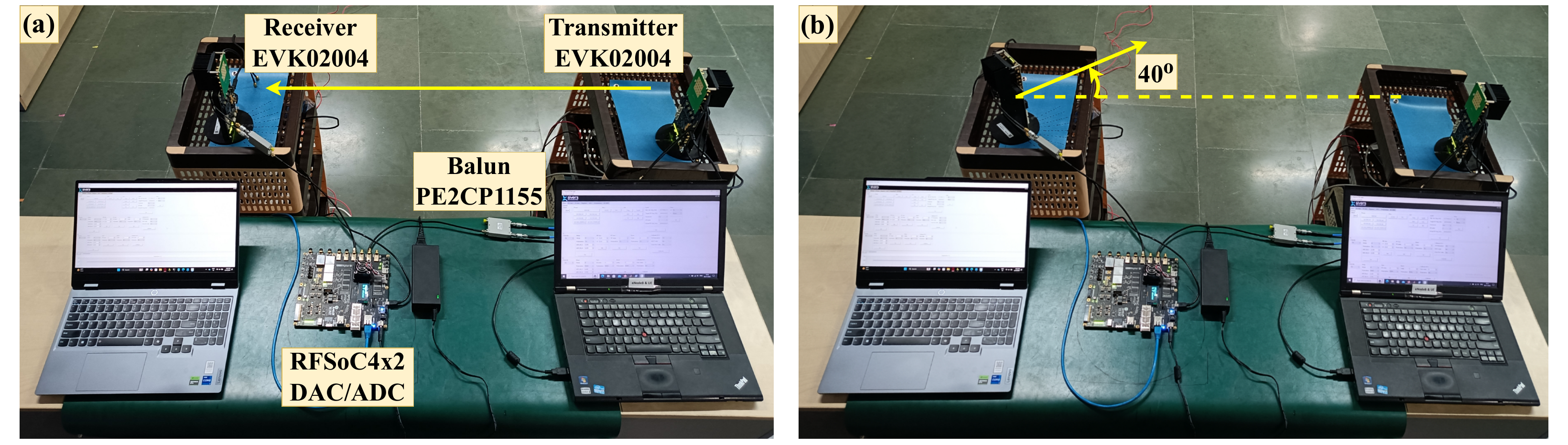}}
\caption{Real view of the proposed hardware testbed when (a) aligned, and (b) misaligned.}
\label{fig:Hardware_Setup}
\end{figure}

\begin{figure}[htbp]
\centerline{\includegraphics[scale=0.3]{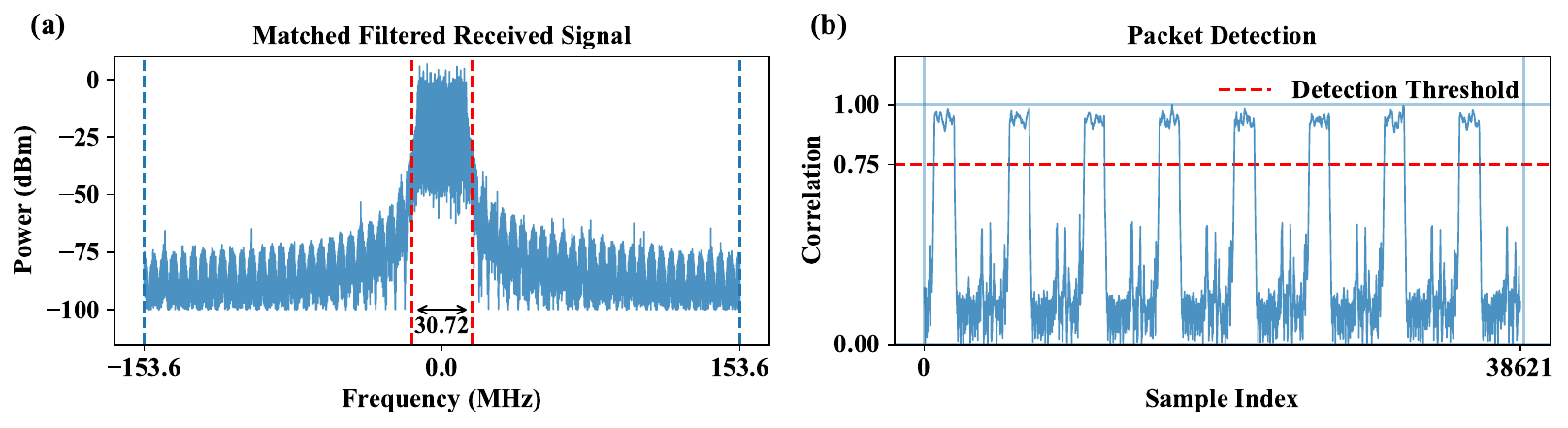}}
\caption{Received OFDM signal when \ac{Tx}-\ac{Rx} are aligned (a) matched filtered output, (b) packet detection}
\label{fig:Rx1}
\end{figure}

\begin{figure}[htbp]
\centerline{\includegraphics[scale=0.3]{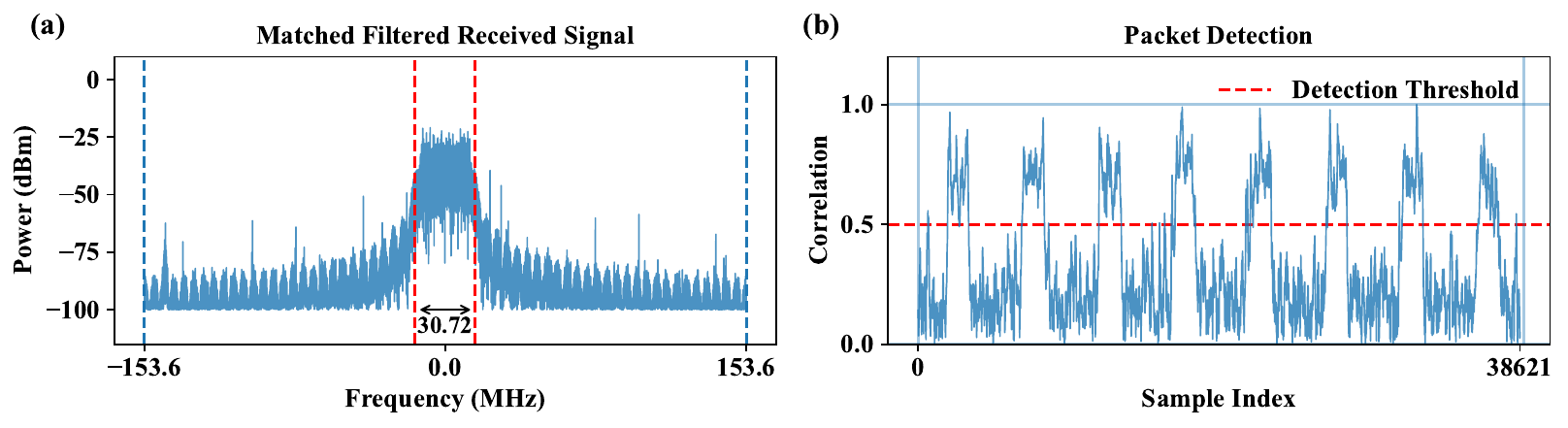}}
\caption{Received OFDM signal when \ac{Tx}-\ac{Rx} are not aligned (a) matched filtered output, (b) packet detection}
\label{fig:Rx2}
\end{figure}

\begin{figure}[htbp]
\centerline{\includegraphics[scale=0.3]{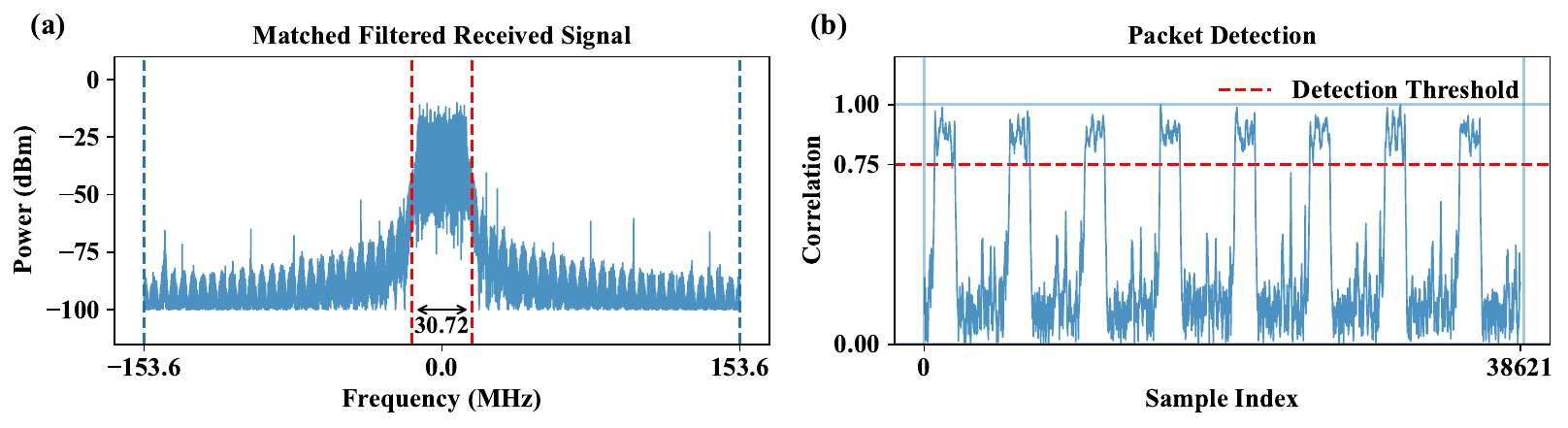}}
\caption{Received OFDM signal when beam of \ac{Tx}-\ac{Rx} are aligned (a) matched filtered output, (b) packet detection}
\label{fig:Rx3}
\end{figure}

We have performed three experiments using the proposed testbed. The first is when antennas of the transmitter and receiver face each other with proper beam alignment. The second is when the receiver antenna is misaligned by $40\degree$ \textcolor{black}{in azimuth} with respect to the transmitter antenna. The third is when the receiver antenna is misaligned $40\degree$ with respect to the transmitter antenna, but the beam is steered with $40\degree$ towards the transmitter using analog beamforming. The hardware setup of the transceiver system is presented in Fig.~\ref{fig:Hardware_Setup}. We have measured the performance of the communication system at \ac{mmW} frequency using \ac{BER} in all three test cases. The received signal for the three test cases are presented in Fig.~\ref{fig:Rx1}, Fig.~\ref{fig:Rx2}, and Fig.~\ref{fig:Rx3}, respectively.

\begin{figure}[htbp]
\centerline{\includegraphics[scale=0.4]{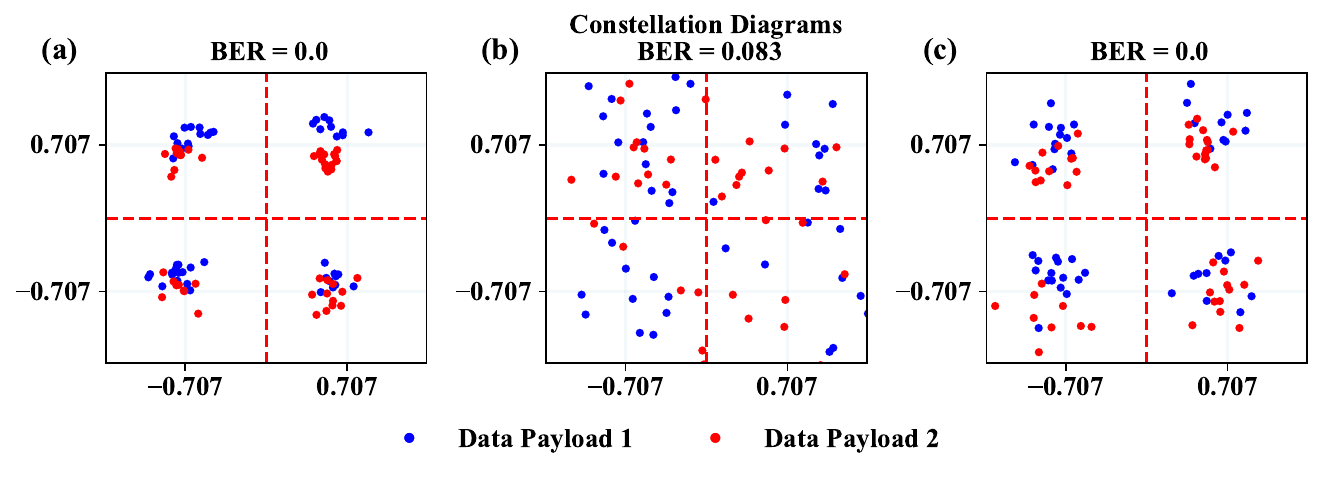}}
\caption{BER of the developed communication testbed (a) \ac{Tx}-\ac{Rx} aligned, (b) \ac{Tx}-\ac{Rx} not aligned, and (c) Beam of \ac{Tx}-\ac{Rx} aligned.}
\label{fig:BER}
\end{figure}

For the first case, the maximum correlation is achieved in packet detection. The packet is detected at a threshold of 0.75 as shown in Fig.~\ref{fig:Rx1}(a). The \ac{BER} is 0 as shown in Fig.~\ref{fig:BER}(a). For the second case, the correlation strength is degraded. However, some amount of signal is still received through the sidelobes and the frame can be detected by lowering the detection threshold to 0.5 as shown in Fig.~\ref{fig:Rx2}(a). The \ac{BER} in this case increases to 0.083 as shown in Fig.~\ref{fig:BER}(b). For the third case, a high correlation is achieved during packet detection as the beams are perfectly aligned using analog beamforming. The packet gets successfully detected at the threshold of 0.75 as shown in Fig.~\ref{fig:Rx3}(a). The \ac{BER} in this case is 0  as shown in Fig.~\ref{fig:BER}(c).

\section{Conclusion \& Future Work}
We have developed and demonstrated a hardware testbed for a wireless transceiver at \ac{mmW} frequency to perform directional communication using analog beamforming. The proposed testbed is designed using AMD RFSoC and the \ac{mmW} analog front-end from Sivers Semiconductor. The performance of the proposed testbed is validated using \ac{BER} for three test cases: (a) when both transmitter and receiver are aligned, (b) when both receiver and transmitter are misaligned, (c) and when both are misaligned but the beams are aligned using analog beamforming. In future work, we will extend the capabilities of the SDR by incorporating the radar functionality for sensing. The communication and radar functionality will then be integrated to demonstrate integrated sensing and communication (ISAC) applications.

\section*{Acknowledgment}
This work is supported by project funds from MeitY, TCS PhD Fellowship (2024-2028), and Qualcomm Innovation Fellowship (2024).

\bibliographystyle{IEEEtran}
\bibliography{ref}

\end{document}